\documentclass[12pt,twoside]{article}   %For printing on two sides of page
\usepackage[super,sort,comma]{natbib}
\usepackage{fancyhdr}		%Gives headers and footers defined below
\usepackage[section]{placeins}   %
\usepackage{graphicx}

\makeatletter \renewcommand\@biblabel[1]{$^{#1}$} \makeatother
\usepackage[mathlines]{lineno}
\usepackage{hyperref}
\hypersetup{ colorlinks,
	citecolor=blue,
	filecolor=blue,
	linkcolor=blue,
	urlcolor=blue
}

\usepackage{xcolor}
\definecolor{gray}{rgb}{0.6,0.6,0.6}
\definecolor{red}{rgb}{0.85,0,0}
\definecolor{green}{rgb}{0,0.85,0}
\definecolor{blue}{rgb}{0,0,0.85}
\definecolor{beige}{rgb}{0.92,0.87,0.78}
\definecolor{amaranth}{rgb}{0.9, 0.17, 0.31}
\usepackage[all]{hypcap}    %causes link to figures to go to figure, not caption
\usepackage{amsmath}
\usepackage{amssymb}
\usepackage[normalem]{ulem}

\begin{document}

% \cen{\sf {\Large {\bfseries Technical Note: adapting the TG-43 formalism for use in Diffusing alpha-emitters Radiation Therapy}} \\  
% 		\vspace*{10mm}
% 		{\Large {\bfseriesGuy Heger, Lior Epstein$^{1,2,3}$, Lior Arazi$^{1,*}$}} \\
	
% {$^{1}$Unit of Nuclear Engineering, Faculty of Engineering Sciences, Ben-Gurion University of the Negav, P.O.B. 653 Be'er-Sheva 8410501, Israel}
% \\ {$^{2}$Department of Biomedical Engineering, Faculty of Engineering, Tel Aviv University, Tel Aviv, Israel}\\
% {$^{3}$Soreq Nuclear Research Center, Yavne, Israel} \\
% 	\vspace{5mm}\\
% 	Version typeset \today\\
% }

\begin{center}
{\Large {\bfseries Adapting the TG-43 formalism for use in Diffusing alpha-emitters Radiation Therapy}}  \\  
\vspace*{10mm}
{\large {\bfseries G. Heger$^{1}$, L. Epstein$^{1,2,3}$ and L. Arazi$^{1,*}$}}\\
\vspace{5mm}
{$^{1}$Unit of Nuclear Engineering, Faculty of Engineering Sciences, Ben-Gurion University of the Negev, Be'er-Sheva, Israel} 
\\ {$^{2}$Department of Biomedical Engineering, Faculty of Engineering, Tel Aviv University, Tel Aviv, Israel}\\
{$^{3}$Soreq Nuclear Research Center, Yavne, Israel} \\

\vspace{3mm}
Version typeset \today\\

\end{center} 

\pagenumbering{arabic}
\setcounter{page}{1}
\pagestyle{plain}
$^*$Corresponding author: Lior Arazi, larazi@bgu.ac.il \\

\begin{abstract}
\noindent 
    {\bf Background:} Diffusing alpha-emitters Radiation Therapy (``Alpha DaRT'') enables the treatment of solid tumors using alpha particles. In Alpha DaRT, the tumor dose distribution is mainly dictated by the diffusion of the decay products of $^{224}$Ra. Due to the inherently different mechanism of dose delivery of the Alpha-DaRT source compared to sources emitting only electrons or photons, the conventional formalism of TG-43 cannot be directly applied to Alpha DaRT.\\
    {\bf Purpose:} To adapt the TG-43 formalism to be used with the Alpha-DaRT model, with the specific aim of allowing commercial treatment-planning software packages to be used for Alpha-DaRT treatment planning.\\
    {\bf Methods:} The \textit{effective} dose rate is defined, along with an activity conversion factor. These, together with a numerical calculation that solves the model equations, allow implementing the Alpha-DaRT model predictions into the TG-43 formalism. \\
    {\bf Results:} An example of calculated $F(r, \theta)$ and $g_L(r)$ tables is presented for an Alpha-DaRT treatment scenario, with the model parameters representing preclinical data on squamous cell carcinoma. \\
    {\bf Conclusions:} Commercial software can be used for Alpha-DaRT treatment planning. However, the generated $F(r, \theta)$ and $g_L(r)$ tables depend on tissue parameters, as well as treatment duration, and so these tables should be generated on a case-by-case basis. 
\end{abstract}
\vspace{2pc}

\newpage
%\maketitle
% \ioptwocol
%

\section{Introduction}
Diffusing alpha-emitters Radiation Therapy (``Alpha DaRT'') enables the treatment of solid tumors using alpha particles \cite{Arazi2007}. The Alpha-DaRT interstitial sources carry a few $\mu$Ci of $^{224}$Ra ($t_{1/2}=3.631$~~d) below their surface. The short-lived decay products of $^{224}$Ra: $^{220}$Rn ($t_{1/2}=55.6$~s), $^{216}$Po ($t_{1/2}=0.144$~s), $^{212}$Pb ($t_{1/2}=10.63$~h), $^{212}$Bi ($t_{1/2}=60.55$~m), $^{212}$Po ($t_{1/2}=0.295~\mu$s) and $^{208}$Tl ($t_{1/2}=3.053$~m)\cite{Nudat3} are released from the source and spread by diffusion (with possible contribution by convective effects), creating---primarily through their alpha decays---a ``kill-region'' measuring several mm in diameter around each source.
The method has been investigated {\it in vitro} \cite{Cooks2009a, Lazarov2011, Cooks2012, Nojima2024} and {\it in vivo} in mice studies as a stand-alone treatment on subcutaneous \cite{Cooks2008, Cooks2009a, Cooks2012} and orthotopic \cite{Zlotnik2026, Cyr2026} tumors, in combination with chemotherapy \cite{Cooks2009b, Horev-Drori2012, Reitkopf-Brodutch2015, Nishri2022}, as an immunostimulator or in combination with immunotherapy \cite{Keisari2014, Confino2015, Confino2016, Domankevich2019, Domankevich2020, Keisari2020, Keisari2021, Del_Mare2023}, and in combination with anti-angiogenic therapy \cite{Nishri2022} and radio-sensitizing drugs \cite{Michaeli2024}. Swine studies served to develop source delivery methods to the lung \cite{Sadoughi2024} and brain \cite{Shoshan2025}.
Clinical trials began in 2017, starting with recurrent or unresectable squamous cell carcinoma (SCC) of the skin, head-and-neck, and oral cavity \cite{Popovtzer2019, Bellia2019, Yang2020, D_Andrea2023, Popovtzer2024}, and expanding to pancreatic tumors \cite{Miller2024}. Ongoing clinical trials are further investigating Alpha DaRT on multiple indications, including brain, prostate, breast, vulvar, lung, and colorectal cancers \cite{AlphaTauWeb}, with clinical practices and radiation safety discussed in \cite{Cohen2024}.
Alpha-DaRT's in-tumor dosimetry has been discussed in both theoretical \cite{Arazi2020, Heger2023a, Heger2023b, Epstein2023, Zhang2024, Cheve2024, Zhang2025, Korotinsky2026, Lielkajis2026} and experimental \cite{Heger2024, Dumancic2025, Cyr2026} publications, with additional studies concerning the dose to distant organs \cite{Arazi2010}, source modeling \cite{Mowlavi2026}, and source characterization \cite{Jollota2025, Deufel2026}.

The theoretical model describing the underlying physics governing the spread of activity inside the tumor during an Alpha-DaRT treatment is called the ``Diffusion-Leakage (DL) model'' \cite{Arazi2020}. The DL equations describe the migration of the three main isotopes in the $^{224}$Ra decay chain: $^{220}$Rn, $^{212}$Pb, and $^{212}$Bi (the others are at local secular equilibrium with their respective parents). The spread of these isotopes was shown to be described by their \textit{diffusion lengths}---$L_{Rn}$, $L_{Pb}$, and $L_{Bi}$---and the $^{212}$Pb leakage probability $P_{leak}(Pb)$ \cite{Arazi2020}. The diffusion lengths are a measure of the average distance traveled by each isotope inside the tumor, from the point of its creation to the point of decay or clearance by the blood. The $^{212}$Pb leakage probability is the probability that a $^{212}$Pb atom released from the source is cleared from the tumor through the blood before its decay. Using the model equations, the dose distribution expected around an Alpha-DaRT source can be calculated by numerically solving the time-dependent DL equations \cite{Heger2023a}. This distribution was shown to depend mainly on the values of the diffusion lengths.

In this work, we show how to translate the calculated dose distribution into the TG-43 formalism \cite{TG43} to allow incorporating the DL model into commercial treatment-planning software (TPS) packages. This requires special considerations, as some of the concepts used for photon/electron-based radiation brachytherapy cannot be immediately applied to Alpha DaRT. 

\section{Adapting the TG-43 formalism to Alpha DaRT}\label{section:tg43-general_considerations}

According to the TG-43 formalism, the dose rate delivered to a point in space is given by:

\begin{equation}\label{eqn:tg43Equation}
    \dot{D}(r,\theta)=S_{K}\cdot \Lambda \cdot \frac{G_L(r,\theta)}{G_L(r_0,\theta_0)}\cdot g_L(r) \cdot F(r,\theta) 
\end{equation}
where $r$ is the radial distance from the source center to the point of interest, and $\theta$ is the angle between the source axis to the line connecting the source center to the point of interest. In this equation:
\begin{itemize}
    \item $S_K$ is the air kerma strength at the reference point $(r_0,\theta_0)$.
    \item $\Lambda$ is the dose-rate constant, defined as $\Lambda \equiv \dot{D}(r_0,\theta_0)/S_K$.
    \item $G_L(r,\theta)$ is the geometry function for a line source approximation, defined as:
        \begin{equation}
            G_L(r,\theta)= \begin{cases}\frac{\beta}{L\,r\,\textrm{sin}(\theta)},& \text{if $\theta\neq 0^\circ$}\\
                                 (r^2-L^2/4)^{-1},              & \text{if $\theta = 0^\circ$} \end{cases}
        \end{equation}
    \item $L$ is the source length.
    \item $\beta$ is is the angle subtended by the tips of the (line) source with respect to the calculation point $(r,\theta)$.
    \item $g_L(r)$ is the radial dose function, defined as
        \begin{equation} \label{eqn:g_eqn}
            g_L(r) = \frac{\dot{D}(r,\theta_0)}{\dot{D}(r_0,\theta_0)} \cdot \frac{G_L(r_0,\theta_0)}{G_L(r,\theta_0)}
        \end{equation}
    \item $F(r,\theta)$ is the anisotropy function, defined as 
        \begin{equation} \label{eqn:F_eqn}
            F(r,\theta) = \frac{\dot{D}(r,\theta)}{\dot{D}(r,\theta_0)} \cdot \frac{G_L(r,\theta_0)}{G_L(r,\theta)}
        \end{equation}
\end{itemize}

To use the TG-43 formalism, one needs to generate $g_L(r)$ and $F(r,\theta)$ tables, as well as determine the values of $S_K$ and $\Lambda$. However, there are several adjustments that need to be taken into account, where the regular TG-43 formalism cannot be applied as-is to Alpha DaRT. The following represents a pragmatic approach, specifically designed to allow the use of commercial brachytherapy TPS packages for Alpha-DaRT treatments:

\begin{itemize}
    \item In conventional low-dose-rate brachytherapy, the dose rate decays exponentially in time according to the decay constant of the source. The total dose at $(r,\theta)$ is the product of the dose rate at that point at $t=0$ and the mean radioactive lifetime of the source. In Alpha DaRT, while the source activity decays with the half-life of $^{224}$Ra, the dose rate outside the source has a complex form that depends on both $r$ and $t$, and is dictated by the diffusion lengths and half-lives of the $^{224}$Ra decay products\cite{Heger2023a}. Moreover, The dose rate outside the source starts at zero and gradually builds up during the treatment, i.e., in an Alpha-DaRT treatment, $\dot{D}(r,\theta,t=0)=0$, which means that Equations \ref{eqn:g_eqn} and \ref{eqn:F_eqn} cannot be used to calculate the values of $g_L(r)$ and $F(r,\theta)$ from the initial dose rate.
    To use the TG-43 formalism, the \textit{effective} dose rate at $t=0$, $\dot{D}_{eff}(r,\theta,t=0)$, is defined using the total dose delivered to a point over the course of a treatment, $D(r,\theta,T)$, with $T$ being the treatment duration. The values of $D(r,\theta,T)$ are obtained by solving the DL equations. Importantly, these values depend on $T$, and hence the effective dose rate at $t=0$ also depends on the treatment duration. Put into equations:

    \begin{equation} \label{eqn:rate_to_total_dose_eqn}
    \begin{split}
        &\dot{D}(r,\theta,t) = \dot{D}_{eff}(r,\theta,0)\cdot e^{-\lambda_{Ra}t} \\
        &D(r,\theta,T) = \int_0^T \dot{D}(r,\theta,t') dt' = \frac{(1-e^{-\lambda_{Ra}T})}{\lambda_{Ra}} \cdot \dot{D}_{eff}(r,\theta,0) \\
        &\dot{D}_{eff}(r,\theta,0) = \frac{\lambda_{Ra}}{(1-e^{-\lambda_{Ra}T})} \cdot D(r,\theta,T)
    \end{split}
    \end{equation}
    Plugging Equation \ref{eqn:rate_to_total_dose_eqn} into Equations \ref{eqn:g_eqn} and \ref{eqn:F_eqn}, it is now possible to use $D(r,\theta,T)$ to calculate $g_L(r,T)$ and $F(r,\theta,T)$, which now depend on $T$ as well:
    \begin{equation} \label{eqn:g_l_F(T)}
    \begin{split}
        &g_L(r,T) = \frac{D(r,\theta_0,T)}{D(r_0,\theta_0,T)} \cdot \frac{G_L(r_0,\theta_0)}{G_L(r,\theta_0)} \\
        &F(r,\theta,T) = \frac{D(r,\theta,T)}{D(r,\theta_0,T)} \cdot \frac{G_L(r,\theta_0)}{G_L(r,\theta)}
    \end{split}
    \end{equation}    
    
    \item The air kerma strength is derived from the photon dose measured in air, typically at 1 m from the source. In an Alpha-DaRT treatment, the dose delivered to the tissue is entirely dependent on the diffusion of the $^{224}$Ra decay products inside the tumor. Since the diffusion coefficients in air are drastically different than those in the tumor volume, measuring the source in air does not represent the dose distribution expected inside the tumor. However, since $S_K$ is proportional to the initial $^{224}$Ra activity on the source $\Gamma_{Ra}(0)$, it can be adapted to the DL model using a conversion factor, $f_{AK}$, so that:
    \begin{equation} \label{eqn:conversio_factor_def}
        S_K = f_{AK} \cdot \Gamma_{Ra}(0)
    \end{equation}
    At the reference point both $g_L(r_0)$ and $F(r_0,\theta_0)$ are equal to 1, so Equation \ref{eqn:tg43Equation} becomes:
    \begin{equation} \label{eqn:rate_at_reference_point}
        \dot{D}_{eff}(r_0,\theta_0,0) = f_{AK} \cdot \Gamma_{Ra}(0) \cdot \Lambda
    \end{equation}
    And the effective dose rate can be converted to total dose using Equation \ref{eqn:rate_to_total_dose_eqn}. Since $f_{AK}$ only defines the relation between $\Gamma_{Ra}(0)$, $S_K$, and $\Lambda$, it can arbitrarily be set to 1, and the values of $S_K$ and $\Lambda$ will change accordingly. Once again, since $\dot{D}_{eff}(r,\theta,T)$ depends on $T$, so does the value of $\Lambda$:
    \begin{equation} \label{eqn:S_k_lambda(T)}
    \begin{split}
        &S_K = \Gamma_{Ra}(0) \\
        &\Lambda(T) = \frac{1}{\Gamma_{Ra}(0)} \cdot \frac{\lambda_{Ra}}{(1-e^{-\lambda_{Ra}T})} \cdot D(r_0,\theta_0,T)
    \end{split}
    \end{equation}     
    
    \item The common reference point ($r=1$ cm, $\theta=90^{\circ}$) is unnatural for an Alpha-DaRT treatment, since at 1~cm from the source the dose is negligible. For that reason, our reference point is set to 2.4 mm from the source (arbitrarily chosen as a point with a therapeutically relevant dose).
\end{itemize}

Based on the above points, the approach used for applying the TG-43 formalism to Alpha DaRT is as follows: calculate the expected dose that will be delivered to a tumor by a single source (for a given source activity and treatment duration), and from the dose map calculate the required tables and constants used in commercial brachytherapy software according to Equations \ref{eqn:g_l_F(T)} and \ref{eqn:S_k_lambda(T)}. The alpha-particle dose $D_{\alpha}$ is calculated by numerically solving the DL equations\cite{Heger2023a}. The low-LET dose by electrons and photons ($D_{\beta+\gamma}$) is calculated using a Monte Carlo code---here, FLUKA \cite{Ahdida2022, Battistoni2015}---and takes into account contributions from both the diffusing isotopes and those remaining on the source \cite{Epstein2023}. Combining the high- and low-LET doses into a single effective dose requires the use of the relative biologic effectiveness (RBE), which, in principle, is tumor-specific, and so separate tables are generated for the high-LET and low-LET dose contributions. In this approach, the RBE-weighted dose is $D_{tot}=RBE\cdot D_{\alpha}+D_{\beta+\gamma}$. In practice, one could pragmatically adopt $RBE=5$ as the value recommended in MIRD Pamphlet 22 for deterministic effects by alpha particles \cite{MIRD_Pamphlet_22}.

\begin{figure}[ht]
\centering
\includegraphics[width=0.25\textwidth]{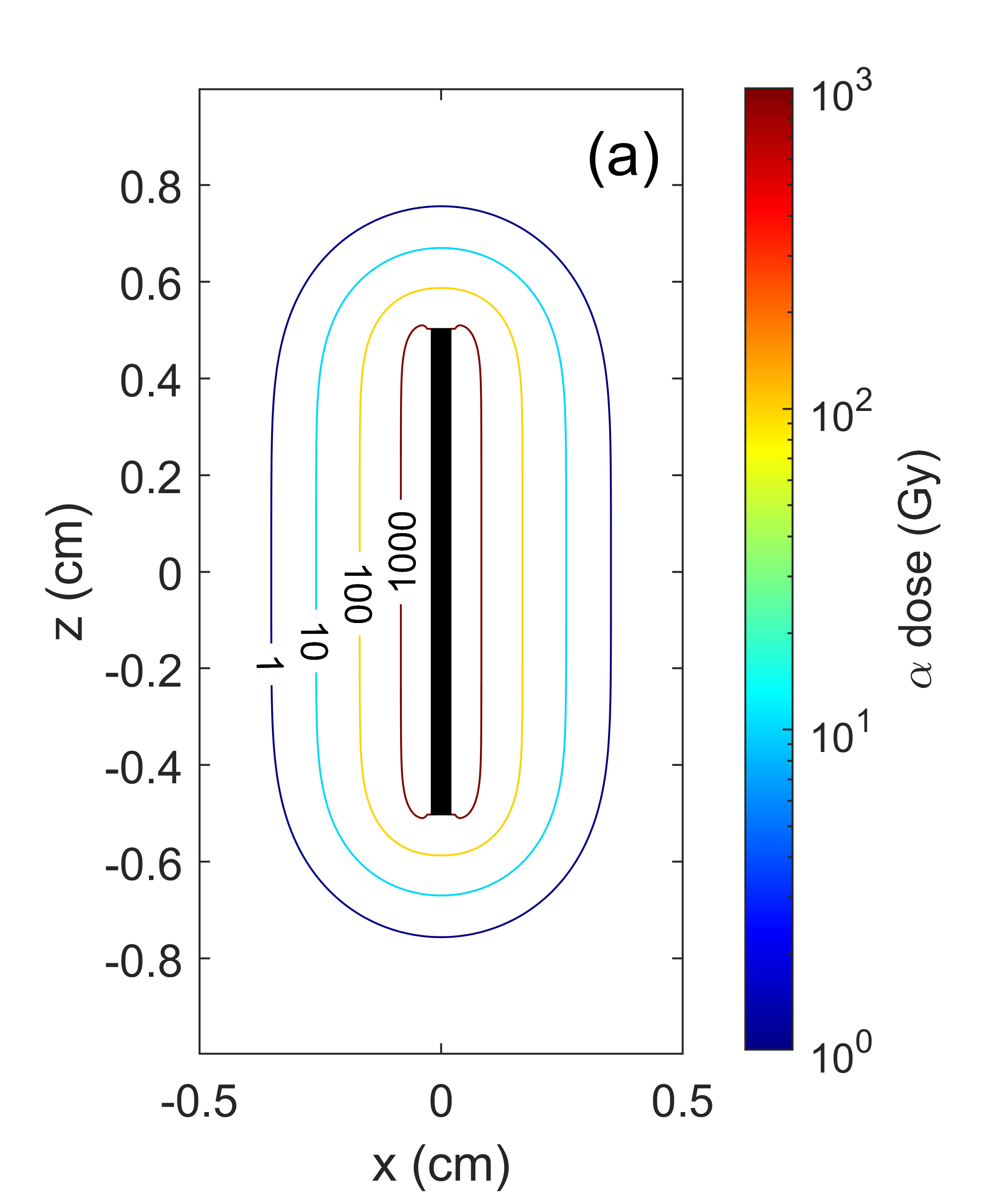}
\includegraphics[width=0.4\textwidth]{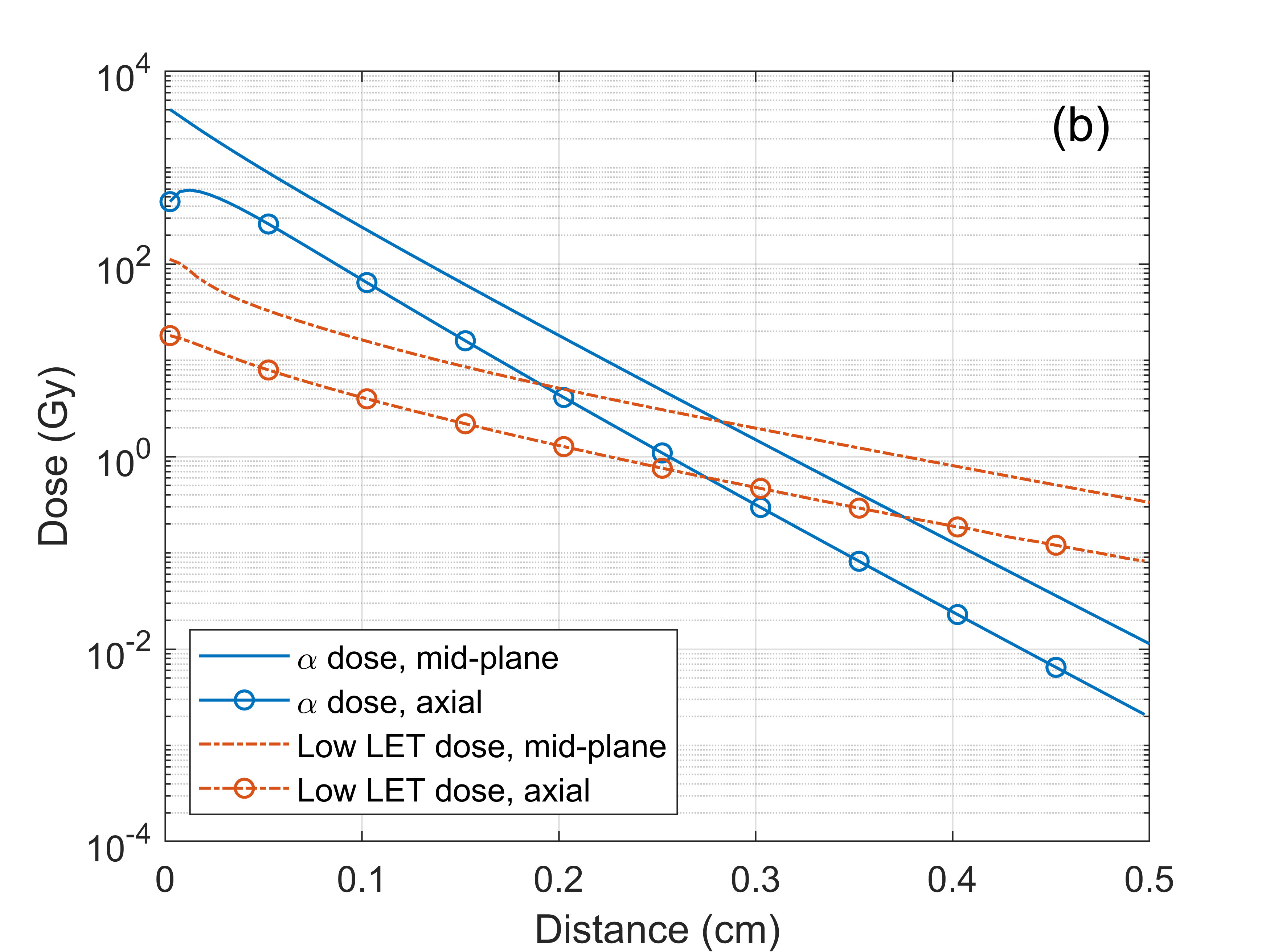}
\includegraphics[width=0.24\textwidth]{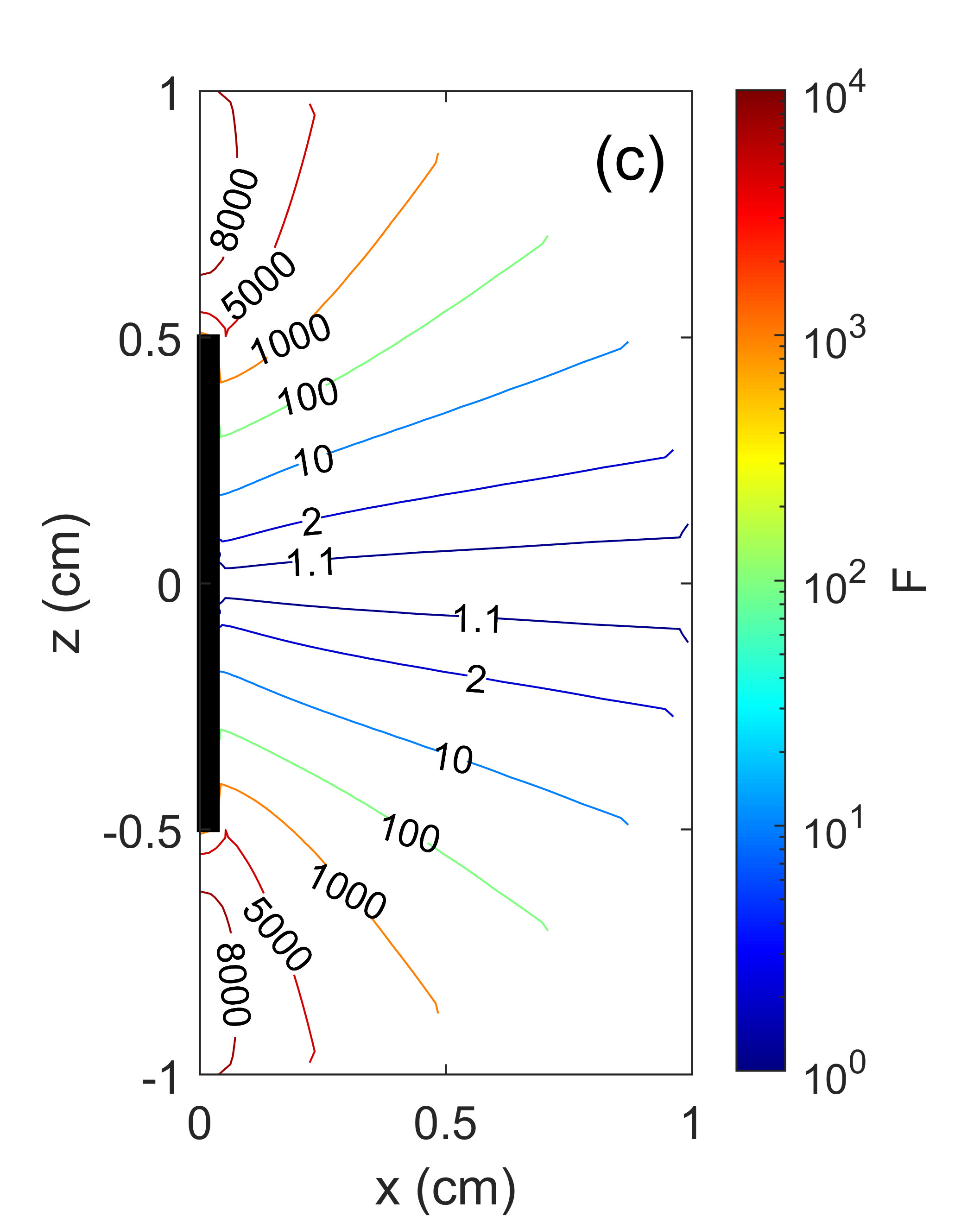}
\caption{(A) Total alpha dose accumulated over 30 days of treatment by a DaRT source with initial activity of $3\;\mu$Ci $^{224}$Ra. The source diameter and length are 0.7\;mm and 10\;mm, respectively. The other model parameters are given in the text. (B) High- and low-LET dose vectors as a function of the distance from the source edge along $r$ in the midplane and along $z$ along the source axis. (C) A graphic representation of calculated $F(r,\theta)$ values based on the model parameters discussed in the text.}
\label{fig:TotalDose2D}
\end{figure}

It is important to note that the dose distribution is specific to each treatment, as the diffusion lengths (and possibly other model parameters) depend on the tumor type \cite{Arazi2020, Heger2024, Dumancic2025}. As an example, one such calculated distribution is displayed in the $xz$ plane (with $z$ being the source axis) in Figure \ref{fig:TotalDose2D}A. The Alpha-DaRT source in this case has a diameter of 0.7~mm and a length of 10~mm. The initial $^{224}$Ra activity of the source is 3\;$\mu$Ci. The diffusion lengths are $L_{Rn} = 0.42$\;mm, $L_{Pb} = 0.2$\;mm, and $L_{Bi} = 0.02$\;mm, representing typical SCC tumors\cite{Dumancic2025}. Additional model parameters\cite{Arazi2020} are: $P_{des}(Rn) = 0.45$, $P_{des}^{eff}(Pb) = 0.55$, $P_{leak}(Pb) = 0.5$, and $\alpha_{Bi}=0$. Figure \ref{fig:TotalDose2D}B shows the radial (at the source mid-plane) and axial (along the source axis) dose profiles as a function of the distance from the source edge, both set such that `0' is the source edge.

\section{Results}
 
The generated $g(r,T)$ and $F(r,\theta,T)$ values for the alpha particle dose, for a treatment period of 30 days, are provided in Table \ref{tab:DaRTFtable} and plotted as a contour plot in Figure \ref{fig:TotalDose2D}C. The calculated values for the low-LET dose are given in Table \ref{tab:DaRTFtable_lowLET}. For comparison, $g(r)$ and $F(r,\theta)$ values for a brachytherapy source with similar geometrical composition containing $^{125}$I (Best Industries model 2301, a low-energy gamma/X-ray emitter \cite{TG43}) are shown in table \ref{tab:BestFtable}. It should be noted that unlike regular $g(r)$ and $F(r,\theta)$ tables, which are typically given for $r$ ranges of up to a few centimeters, for the Alpha-DaRT tables, the distances are given in the range 0.1 - 0.6 cm, since the dose in greater distances is negligible (see Figure \ref{fig:TotalDose2D}B).
As can be seen in Table \ref{tab:DaRTFtable}, for the $\alpha$ dose, the values of $F(r,\theta)$ change over several orders of magnitude -- a drastic change when compared to photon-based brachytherapy tables. This is due to the rapid fall-off of the $\alpha$ dose around the DaRT source, as can be seen in Figure \ref{fig:TotalDose2D}A. On the other hand, for the low-LET dose, shown in table \ref{tab:DaRTFtable_lowLET}, the change is much smaller and is more akin to the values for commercial low-energy gamma/X-ray sources. 

\begin{table}[]
\caption{Calculated values of $F(r,\theta,T)$ and $g_L(r)$ for an Alpha-DaRT source's alpha dose (treatment period of 30 days, $L_{Pb} = 0.2$\;mm, $L_{Rn} = 0.42$\;mm and $L_{Bi} = 0.02$\;mm). Values of $F(r,\theta,T)$ for coordinates that lie within the source dimensions are marked with '---'. \\}
\centering
    % \begin{tabular}{ | p{1cm} | p{1cm} | p{1cm} | p{1cm} | p{1cm} | p{1cm} | p{1cm} | p{1cm} |}
    \begin{tabular}{ | c | c  c  c  c  c  c |}
    \hline
    \multicolumn{7}{|c|}{$F(r,\theta)$ - DaRT source} \\
    \hline
    % \multirow{2}{4em}
     & \multicolumn{6}{c|}{r (cm)} \\
    $\theta$ (deg) & 0.1 & 0.2 & 0.3 & 0.4 & 0.5 & 0.6\\
    \hline
    0  & ---   &    ---   & 	---   & 	---   &     ---   &     9.5E3\\
    10 & ---   &    ---   &     102.4 &     718.0 &     4.0E3 &     6.7E3\\
    20 & ---   &    10.75 &     49.68 &     237.4 &     926.2 &     2.0E3\\
    30 & 2.071 & 	6.576 &	    21.24 &     69.87 &     201.3 &     421.2\\
    40 & 1.740 &	3.976 &	    9.239 &	    21.70 &	    47.43 &     87.80\\
    50 & 1.458 &	2.508 &	    4.368 &	    7.665 &	    13.03 &	    20.48\\
    60 & 1.248 &	1.703 &	    2.343 &	    3.234 &	    4.429 &	    5.877\\
    70 & 1.107 &	1.273 &	    1.470 &	    1.702 &	    1.965 &	    2.250\\
    80 & 1.026 &	1.063 &	    1.102 &	    1.144 &	    1.187 &	    1.230\\
    \hline
    $g_L(r)$ & 14.36 & 2.223 & 0.293 & 0.036 & 4.3E-3 & 4.9E-4\\
    \hline
    \end{tabular}
\label{tab:DaRTFtable}
\end{table}

\begin{table}[]
\caption{Calculated values of $F(r,\theta,T)$ and $g_L(r)$ for an Alpha-DaRT source's low-LET dose (treatment period of 30 days, $L_{Pb} = 0.2$\;mm, $L_{Rn} = 0.42$\;mm and $L_{Bi} = 0.02$\;mm). Values of $F(r,\theta,T)$ for coordinates that lie within the source dimensions are marked with '---'. \\}
\centering
    % \begin{tabular}{ | p{1cm} | p{1cm} | p{1cm} | p{1cm} | p{1cm} | p{1cm} | p{1cm} | p{1cm} |}
    \begin{tabular}{ | c | c  c  c  c  c  c |}
    \hline
    \multicolumn{7}{|c|}{$F(r,\theta)$ - DaRT source} \\
    \hline
    % \multirow{2}{4em}
     & \multicolumn{6}{c|}{r (cm)} \\
    $\theta$ (deg) & 0.1 & 0.2 & 0.3 & 0.4 & 0.5 & 0.6\\
    \hline
    0  & ---   &    ---   & 	---   & 	---   &     ---   &     8.144\\
    10 & ---   &    ---   &     2.554 &     3.743 &     5.611 &     7.061\\
    20 & ---   &    1.579 &     2.103 &     2.956 &     4.131 &     5.330\\
    30 & 1.222 & 	1.418 &	    1.788 &     2.354 &     3.078 &     3.855\\
    40 & 1.136 &	1.286 &	    1.537 &	    1.885 &	    2.310 &     2.764\\
    50 & 1.083 &	1.183 &	    1.336 &	    1.535 &	    1.768 &	    2.011\\
    60 & 1.046 &	1.103 &	    1.185 &	    1.287 &	    1.401 &	    1.518\\
    70 & 1.022 &	1.046 &	    1.081 &	    1.123 &	    1.169 &	    1.213\\
    80 & 1.005 &	1.011 &	    1.020 &	    1.030 &	    1.041 &	    1.051\\
    \hline
    $g_L(r)$ & 1.635 & 1.161 & 0.784 & 0.495 & 0.293 & 0.163\\
    \hline
    \end{tabular}
\label{tab:DaRTFtable_lowLET}
\end{table}

\begin{table}[]
\caption{Cited values of $F(r,\theta)$ and $g_L(r)$ for a low-energy gamma/X-ray-emitting brachytherapy source (Best Industries model 2301)\cite{TG43}. \\}
\centering
    \begin{tabular}{| c | c  c  c  c  c  c |}
    \hline
    \multicolumn{7}{|c|}{$F(r,\theta)$ - 'Best' model 2301 source} \\
    \hline
    & \multicolumn{6}{c|}{r (cm)} \\
    $\theta$ (deg)  & 1 & 2 & 3 & 4 & 5 & 6\\
    \hline
    0  & 0.367 & 0.454 & 0.922 & 0.902 & 0.894 & 0.893\\
    10 & 0.653 & 0.671 & 0.699 & 0.727 & 0.732 & 0.764\\
    20 & 0.785 & 0.794 & 0.809 & 0.814 & 0.825 & 0.852\\
    30 & 0.900 & 0.890 & 0.885 & 0.892 & 0.899 & 0.915\\
    40 & 0.982 & 0.954 & 0.947 & 0.939 & 0.943 & 0.976\\
    50 & 1.014 & 0.992 & 0.985 & 0.991 & 0.997 & 0.989\\
    60 & 1.030 & 1.010 & 1.009 & 1.007 & 1.010 & 1.019\\
    70 & 1.036 & 1.026 & 1.016 & 1.023 & 1.011 & 1.035\\
    80 & 1.010 & 1.030 & 1.019 & 1.017 & 1.010 & 1.020\\
    \hline
    $g_L(r)$ & 1.000 & 0.866 & 0.707 & 0.555 & 0.427 & 0.320\\
    \hline
    \end{tabular}
\label{tab:BestFtable} 
\end{table}

\clearpage

\section{Discussion}

In this paper, a practical approach was presented for adapting the TG-43 formalism to the DL model, thus allowing the use of commercial brachytherapy software for Alpha-DaRT treatment planning. Due to the inherently different behavior of alpha particles compared to beta/gamma, some modifications to the common practice were applied: since the dose rate outside the source starts at 0 and gradually builds up during the treatment, the \textit{effective} dose rate at $t=0$ was introduced; A new conversion factor, $f_{AK}$, was also introduced, which accounts for the fact that measuring the photon dose 1~m from the source, in air, is irrelevant for an Alpha-DaRT treatment. However, since $S_K$ is proportional to the initial $^{224}$Ra activity in the source, a conversion factor can be used to adapt it to the DL model; Finally, the reference point is set to 2.4~mm from the source surface, since at 1~cm from the source the dose is negligible.\\

It is important to note the drastic change in the values of $F(r, \theta)$ calculated for an Alpha-DaRT treatment (Table \ref{tab:DaRTFtable}) compared to those of conventional brachytherapy table (\ref{tab:BestFtable}): while the latter is relatively constant and remains on the scale of 1 over radii of up 6 cm, the former changes by 3 orders of magnitude over the relevant distance of 0.6 cm. This is highlighted in figure \ref{fig:TotalDose2D}C, where a contour map of $F(r, \theta)$ values is shown. The same is also true for $g_L(r)$. For that reason, when calculating TG-43 tables for DaRT, a much finer resolution is required -- order of 0.1-0.2 mm in $r$ and 1$^\circ$-2$^\circ$ in $\theta$, compared to conventional tables.

\section*{Acknowledgments}
The authors wish to thank Dr. Christopher Deufel for his thorough reading and thoughtful comments on this manuscript.

\section*{Conflict of Interests}
This work was partly funded by Alpha TAU Medical Ltd (ATM). L.A. is a minor shareholder of ATM. The scholarship of G.H. was partially paid by ATM through a research agreement with Ben-Gurion University of the Negev. L.A. is co-inventor of several DaRT-related patents. G.H. is a co-inventor of a pending patent application on DaRT dose calculations.

\section*{References}
\bibliography{DaRT_alpha_modeling_bib}
\bibliographystyle{./medphy.bst}

\end{document}